\documentstyle[twocolumn,aps,prl,psfig]{revtex}
\begin{document}
\draft
\bibliographystyle{unsrt}
\title {STOCHASTIC PIONISATION IN HOT QUARK-GLUON MATTER}
\vskip 36pt
\author{Jan -e Alam$^{1,a}$,  Abhijit Bhattacharyya$^{1,b}$
, Sanjay K. Ghosh$^{2,c}$,
Sibaji Raha$^{2,d}$
 and  Bikash Sinha$^{1,3,e}$}
\vskip 20pt

\address{$^1$ Variable Energy Cyclotron Centre, 1/AF, Bidhannagar,\\ 
Calcutta- 700 064, India\\
$^2$Department of Physics, Bose Institute, 
93/1, A.P.C.Road,\\
Calcutta 700 009, India \\
$^3$Saha Institute of Nuclear Physics, 1/AF, Bidhannagar, \\
Calcutta- 700 064, India.}
\vskip 24pt
\maketitle
\begin{abstract}
We present a microscopic approach to dynamical pionisation of the hot 
quark-gluon matter formed in ultrarelativistic heavy ion collisions. The 
time evolution of the system is described assuming that quarks undergo 
Brownian motion in a thermal bath provided by the gluons. The rate of 
hadronization as well as the time dependence of the temperature of the
system are seen to be quite sensitive to the QCD $\Lambda$ parameter. Even 
in a non-equilibrium scenario, we find that 
there appears a
clear hint of a first order phase transition.
\end{abstract}
\vskip 10pt
\par
Search for the novel state of strongly interacting matter called 
Quark-Gluon Plasma (QGP), in which the properties of bulk matter are
determined by the quarks and gluons rather than the usual hadrons, is a
major sphere of current activity. Quantum Chromodynamics (QCD) predicts
that such a phase, {\it i.e.}, a metastable configuration, should exist
at very high temperature and/or baryon density. It is strongly believed that
conditions conducive to the formation of QGP could indeed be achieved, 
albeit transiently, in highly energetic collisions of heavy ions and
consequently, a substantial effort is now being devoted to look for QGP
in such collisions.
\par
An inescapable feature of the collision process is that, because of 
colour confinement, the quarks and 
gluons in the QGP must, at some epoch, turn into hadrons which would 
ultimately be detected, never the individual quarks and gluons. An {\it ab
initio} calculation of this very important, non-perturbative, phenomenon is, 
unfortunately, beyond the capability of QCD as yet. It has thus been widely 
postulated that this conversion of coloured particles to the colourless 
hadrons en bulk could be an actual phase transition (the order of which is 
an open issue) \cite{a}. Assuming the phase transition to be of first order,
some authors \cite{b} have attempted to investigate the process of
hadronization in a QGP through a bubble nucleation mechanism. In addition
to being a classical picture, this mechanism also requires thermodynamic
equilibrium. Recent studies on QGP dynamics \cite{d,e,f}, showing the lack of 
thermodynamic equilibrium in the quark-gluon phase in ultrarelativistic 
heavy ion collisions, indicate that such an ideal situation is rather 
unlikely. 
\par
Recently there have been some attempts to study the formation of hadrons
in quark matter using different semi-microscopic approaches. These studies 
can be characterized either as model dependent calculations \cite{reberg},
or the computer codes based on the string phenomenology \cite{string}, or
other phenomenological description of hadronization \cite{barz}. None
of these approaches account for the essential lack of equilibrium in the
quark-gluon phase. Some efforts have also been made to estimate hadronization
within the parton cascade model by introducing a cut-off to mimic the 
non-perturbative effects \cite{ellis}. To the best of our knowledge, the first study aimed at
investigating the dynamical process of hadronization in a non-equilibrated
quark-gluon system from a physically transparent approach was in \cite{kyoto}. 
In this letter, we present the 
details of this scenario describing the evolution of the non-equilibrated 
hot quark-gluon matter towards hadronization. For simplicity, we have 
restricted ourselves to the case of pions only in the present work; since
the pions account for the bulk of the multiplicity, the conclusions should
remain valid not only qualitatively and even semi-quantitatively. 
\par
We have earlier shown \cite{d,f} that perturbative estimates of the 
gluon-gluon, quark-gluon and quark-quark cross sections allow us to 
study the evolution of the quark-gluon matter formed in ultrarelativistic
heavy ion collisions by visualizing the quarks as Brownian particles in a 
hot gluonic thermal bath. In this work, we start with such a premise, which,
in our opinion, describes the non-equilibrium aspects of the evolution in a
physically transparent manner.
\par
Hadronization in such a system can be studied in the light of Smoluchowski's
theory of coagulation in colloids, which was elaborated further by
Chandrasekhar \cite{chandra}. This theory suggests that coagulation 
results as a consequence of each colloidal particle being surrounded by
a sphere of influence of a certain radius $R$ such that the Brownian 
motion of a particle proceeds unaffected only so long as no other particle
comes within its sphere of influence. When another Brownian particle does 
come within a distance $R$ of the test particle, they form a two body cluster. 
This cluster also describes a Brownian motion but at a reduced rate due to 
its increased size/mass. The process continues further till a single cluster 
of all the particles is formed. 
\par 
In order for this stochastic scenario of cluster formation to apply 
to a system of Brownian quarks in the 
hot gluon bath, it is essential that each quark has an appropriate sphere of 
influence of radius $r$. Obviously, this radius $r$ will depend on the 
spin-isospin combination of final cluster (whether the final cluster is 
scalar, pseudoscalar, vector or axial vector meson or even a baryon).
Mesons are formed when one quark and one antiquark with proper quantum numbers
come within the spheres of influence of each other. (Clusters with 
greater numbers of quarks and antiquarks (e.g. baryons) can also be formed by 
imposing the conditions of colour neutrality and charge balance properly.) 
This implies that the radius of the sphere of influence corresponds to the 
correlation length between the quarks in the proper hadronic channel. In 
other words, this is the screening length of the corresponding hadrons in 
the hot quark-gluon matter. For the present purpose, the radius corresponds
to the screening length of pions, which can be evaluated in, say, the 
Nambu-Jona-Lasinio (NJL) model \cite{florkow} or the lattice \cite{lattice}. 
(One can immediately see, without further ado, that the rate of pion formation
in the hot quark matter should be rather small at high temperatures and 
increase with falling temperature as the pion screening length (inverse of 
the screening mass) decreases with increasing temperature in all realistic 
pictures.) The pions formed at very high temperatures are most likely to 
decay back into quarks and antiquarks. It has been recently shown, within 
the NJL model, that the pion decay width is very large at high temperatures 
and becomes zero around a temperature of $215$ MeV or so \cite{abhijit}. 
Since this would be a crucial ingredient in what follows, we adopt, for the 
sake of consistency, the NJL model estimates for the pion screening mass in 
the present work. Note that evaluating the pion decay width as a function of
temperature on the lattice is not possible as the dynamical pion mass 
cannot be evaluated there.
\par
The rate of pion formation from Brownian quarks as a stochastic 
process, as is evident from the preceding discussion, 
depends on the number of quarks (antiquarks) falling into the sphere of
influence of another antiquark (quark), the number of pions decaying back to 
quarks and antiquarks and also the change in the pion density due to the 
expansion of the system. Thus the rate equations can be written as,
\begin{eqnarray}
{dn_{\pi^{a}} \over dt}= n_{q_i} n_{\bar{q}_j} 4\pi r^2<\vec{v}\cdot \hat{r}>
\nonumber\\ 
- {\Gamma^{total}}_{\pi^{a} \rightarrow q_{i} {\bar q}_{j}} - {n_{\pi^a}\over t}  
 \label{eq:dp1dt} \\
{dn_{\pi^{0}} \over dt}= {1\over 2} (n_{u} n_{\bar{u}}+ n_{d} n_{\bar{d}})     
4\pi r^2<\vec{v}\cdot \hat{r}> \nonumber\\
-{\Gamma^{total}}_{{\pi^0} \rightarrow u \bar{u}
(d{\bar d})} - {n_{\pi^0}\over t}  {\label{eq:dp2dt}}\\
{dn_{q_{i}} \over dt} =  {\Gamma^{total}}_{g\rightarrow q_{i} \bar{q}_{i}}
+{\Gamma^{total}}_{gg\rightarrow q_{i} \bar{q}_{i}}  \nonumber\\  
- n_{q_i} n_{\bar{q}_j} 4\pi r^2 <\vec{v}\cdot \hat{r}>+ 
{\Gamma^{total}}_{\pi \rightarrow q_{i} {\bar q}_{j}} - {n_{q_i}\over t}
{\label{eq:dqdt}}  
\end{eqnarray}

In eqs.(\ref{eq:dp1dt},\ref{eq:dp2dt}), the first term is the rate of 
pion formation 
($a\equiv$ + or -); the second term is the rate of pions decaying back to 
quarks and the third term is due to  Bjorken (longitudinal) expansion of 
the system. $i(j)$ stands for $u$ or $d$ (we ignore $s$ and other heavier 
flavours). $<\vec {v} \cdot \hat{r}>$ (the average 
relative velocity in the radial direction) is calculated using the 
J\"uttner distribution,
\begin{equation}
f(x,p)=e^{-\beta p \cdot u(x)} \label{eq:jutt}
\end{equation}
There would also be a corresponding rate equation for the antiquarks, which
looks exactly like eq. (\ref{eq:dqdt}) and hence not explicitly written.
In eq. (\ref{eq:dqdt}) the ${\Gamma^{total}}_{g\rightarrow q \bar q}$ as well
as ${\Gamma^{total}}_{gg\rightarrow q \bar q}$ stand for the corresponding
net quantities.
\par
As mentioned earlier, we are considering a non-equilibrated quark matter
and hence the pions formed will also be out of equilibrium. This is
taken into account by multiplying the relevant distribution functions with 
the  ratios $r_{q}=n_{q}/n_{eq}$ and $r_{\pi}=n_{\pi}/n_{e\pi}$ where
$n_{q}$ and $n_{eq}$ are non-equilibrium and equilibrium densities of
quarks and $n_{\pi}$ and $n_{e\pi}$ are non-equilibrium and equilibrium
densities for pions. These details are given in \cite{f}.
\par
In all these expressions, the appropriate masses are the effective masses 
including the current as well as thermal contribution,
, whose importance in determining the dynamics of the hot quark matter
has been well established. For quarks (antiquarks), this is 
$$m_{eff}=\sqrt{{m_{q}(curr)}^{2}+{m_{q}(thermal)}^{2}}$$ 
where \cite{alther,thoma}
\begin{equation}
{m_{q}^{2}}(thermal)= (1+{r_{q}\over {2}}) ({{g_{s}T} \over 3})^{2}
\end{equation}
and $m_{q}(curr)$ is taken to be $10$ MeV. For gluons the thermal mass is,
\begin{equation}
m_{g}(thermal)={2\over 3} g_{s}T
\end{equation}
The running coupling constant $\alpha_{s}$ as a function of temperature is 
given by \cite{geiger}
\begin{equation}
\alpha_{s}={{12\pi} \over {(33-2n_{f})ln \left[ {\bar{Q^2} \over \Lambda^2} \right]}}
\end{equation}
with $\bar{Q^2} = {m_{eff}}^{2}(T) + 9T^{2}$. 
\par
Simultaneously, we must take account of energy momentum conservation
which, for a Bjorken flow, corresponds to the following equation
\begin{equation}
{\partial{\epsilon} \over {\partial t}}= - {{\epsilon + P} \over t}
\label{eq:energy}
\end{equation}
where $\epsilon\equiv \epsilon_{total} = \epsilon_{g} + \epsilon_{q}
+ \epsilon_{\pi}$. We also include the one loop correction to
$\epsilon_{g}$ \cite{plumer}. $\epsilon$ and $P$ are related through
the velocity of sound, as in \cite{f}. For a complete description of 
the system, eqs. ({\ref{eq:dp1dt}), (\ref{eq:dp2dt}), (\ref{eq:dqdt}) 
and (\ref{eq:energy}) must be solved self-consistently. The initial 
conditions are taken from \cite{f} for RHIC energies. The initial 
time ($t_{g}$) is the time when gluons thermalise (=0.3 fm), where 
$r_{q}$ = 0.15, $r_{g}$=1 and $r_{\pi}$ is taken to be 0. The 
temperature at this time is 500 MeV. The pion decay width, dynamical 
mass and screening mass are taken, as already mentioned, from the 
NJL model \cite{florkow,abhijit}. Note that we are working at $y=0$ 
so that $t$ and $\tau$ are the same 
and the baryon chemical potential is zero.
\par
\begin{figure}[htb]
\psfig{file=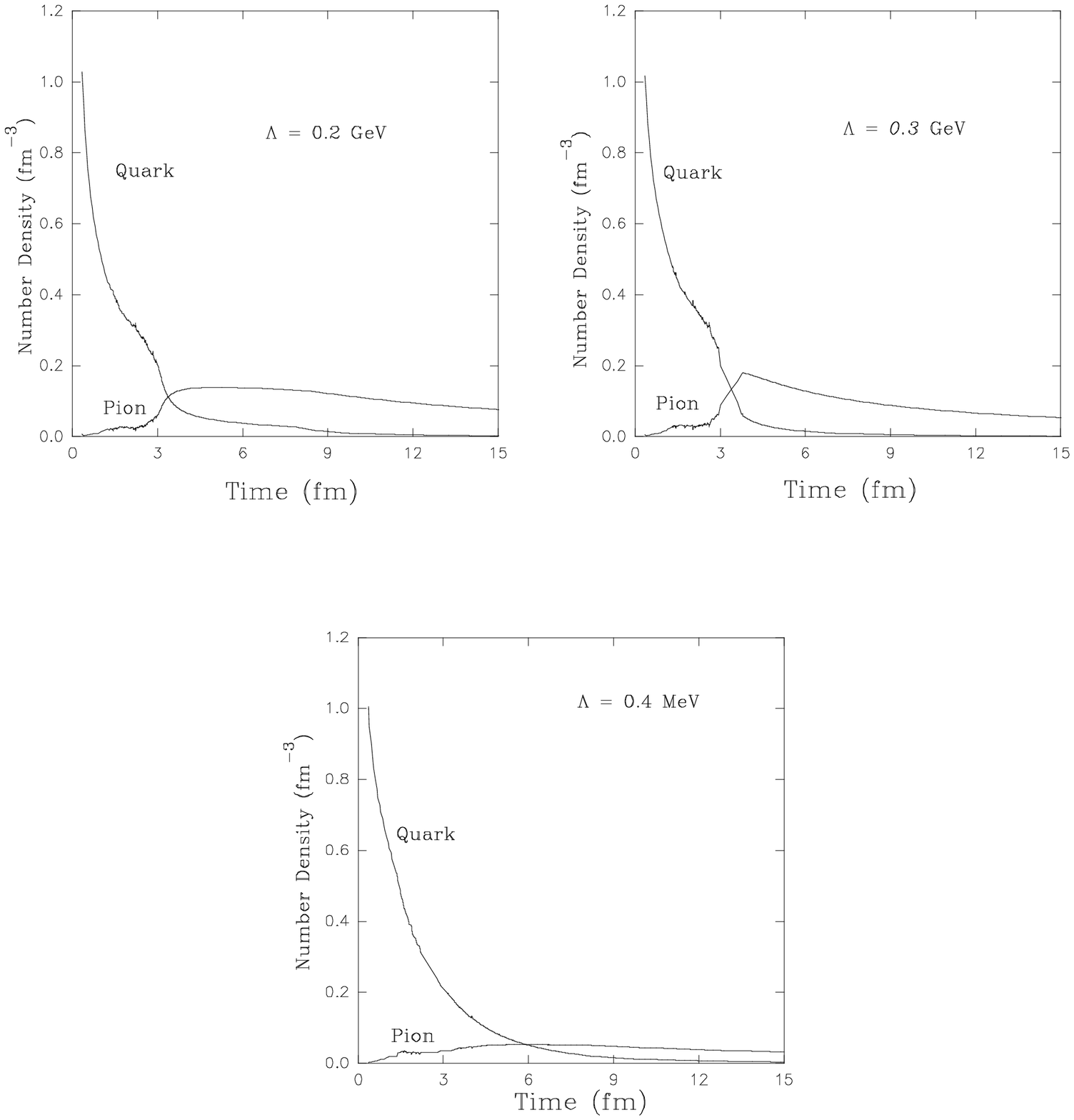,width=3.3in,height=5in}
\end{figure}
\vskip 10pt
\begin{center}
Figure 1 : Time evolution of quark and pion number densities for various
values of QCD parameter $\Lambda$.
\end{center}
\par 
Figure1 shows the variation of pion and quark number densities with 
time, for various values of the QCD parameter $\Lambda$. In all three cases, 
we find the same qualitative feature that pions start appearing in the system
quite early on but they become appreciable in number only after some time.
At late times the system is dominated by pions. This cross over occurs 
at $t \geq $ 4 fm for $\Lambda$ = 0.2 or 0.3 GeV while for $\Lambda$=
0.4 GeV this happens at $t\sim$ 6 fm. 
\par
Figure 2 shows the variation of temperature with time. Obviously, there is
a dramatic effect of the QCD parameter $\Lambda$. In all the cases, there
is a change at $T\sim$ 215 MeV, corresponding to $t\sim$ 3.5 fm; the 
variation of temperature with time becomes slower, as is expected in the
mixed phase of a first order phase transition. At $\Lambda$=0.2 GeV,
this occurs for a very short period of time, before the system starts cooling
again. The duration of the constant temperature configuration  increases
with $\Lambda$, and for $\Lambda$=0.4 GeV, it persist upto 9 fm before the
temperature of the system starts falling again. 
\par
\begin{figure}[htb]
\psfig{file=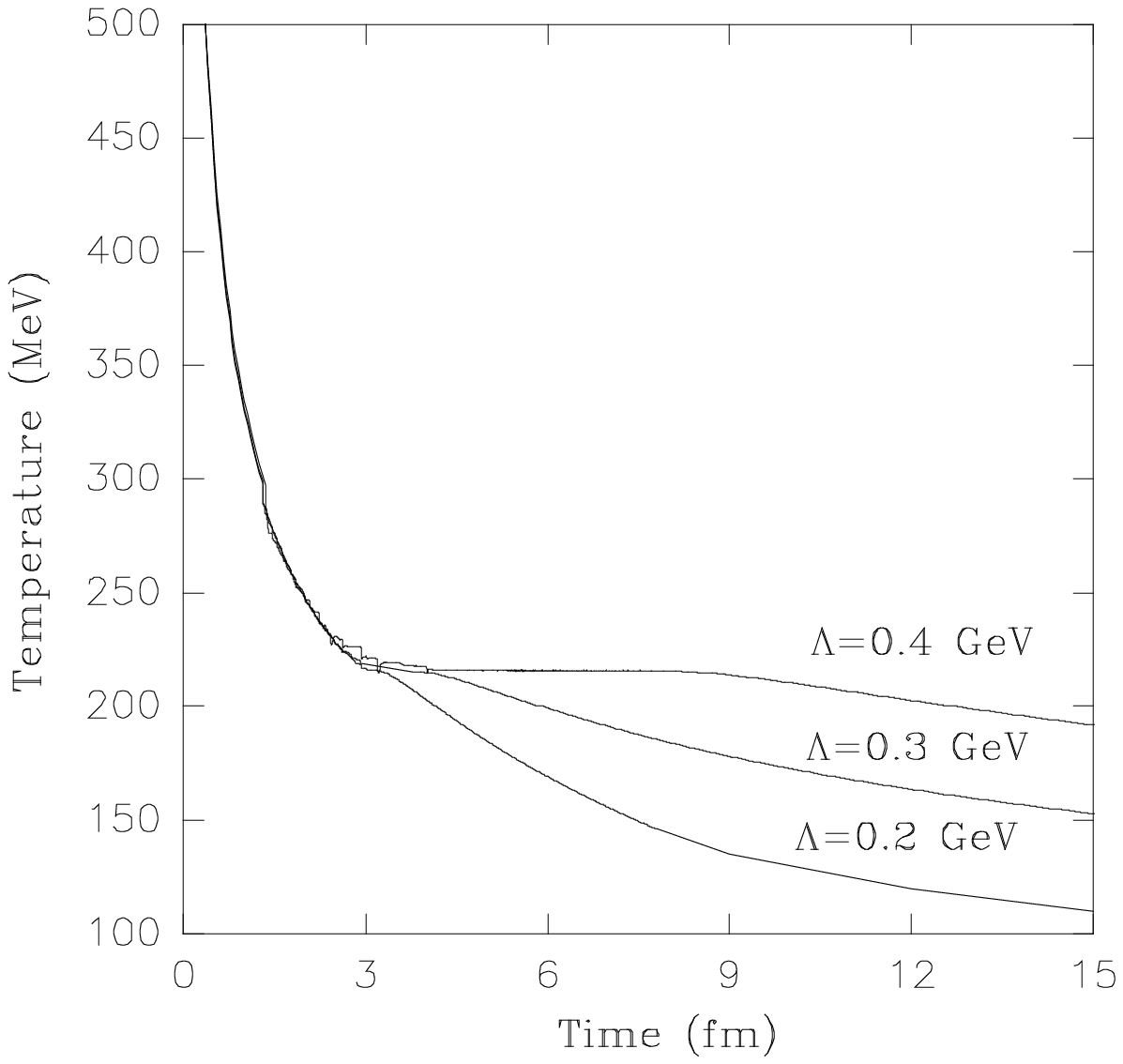,width=3in,height=3in}
\end{figure}
\vskip 10pt
\begin{center}
Figure 2 : Time evolution of the Temperature for various values of 
QCD parameter $\Lambda$.
\end{center}
\par
Obviously, this is a clear indication of an {\it apparent} first order
transition. Microscopically, the appearance of the mixed phase at
a temperature of $\sim$ 215 MeV can be understood from the fact that
the pion decay width goes to zero at such a temperature \cite{abhijit}.
All the pions that were formed earlier in the system tended to decay back
to quarks and antiquarks on a fast time scale. Only after the pion decay
width becomes small would the formed pions become stable. 
\par
The role of the  QCD parameter $\Lambda$ is better understood from 
figure 2. The higher the value of $\Lambda$, the higher the quark
thermal masses. As a result, the lower the relative velocity, which
would lead to a lower rate of pion production as well as a slower
depletion of the quark number density. Thus the mixed phase, 
{\it i.e.} the domain where quarks and pions have comparable 
densities, would not only occur later in time but also persist for 
a longer duration. It should however be mentioned that there are 
also other effects associated with the QCD $\Lambda$ (like the quark 
production rate from gluon fusion and /or decay, the gluon energy 
density and so on) which have competing roles in determining the 
actual number densitites. This may be the reason why 
one does not notice a drastic difference between $\Lambda$ = 0.2 and
0.3 GeV in figure 1, while at $\Lambda$ = 0.4 GeV, the change is 
more noticeable even in figure 1. A detailed analysis of all these 
different effects is in progress now.  
\par
We must stress the fact that the first order phase transition is only 
an approximate one. There is no unique demarcation between the different
phases. High temperature domains are dominated by quarks with a few pions 
present in the system, the intermediate region ($T\sim$ 215 MeV) having
comparable numbers of quarks and pions and the low temperature domain 
being mostly pions with a few quarks. Nonetheless, it can be fairly 
concluded from these results that the concept of a first order confining 
phase transition is not a bad approximation for the process of
hadronization in QGP. It should also be noted that the so called 
critical temperature of $\sim$ 215 MeV, which corresponds to the temperature
where the pion decay width vanishes, is a model dependent quantity.
In our case this derives from the NJL model, but the qualitative features
should of course be model independent.
\par
In conclusion, we have studied, in a physically transparent picture, 
the dynamical process of hadronization (pion formation) in a
non-equilibrated quark-gluon system formed in ultrarelativistic 
heavy ion collision. Our results show that even in a microscopic analysis,
there is a clear indication of an apparent first order phase transition
in the system. Nonetheless, the persistence of non-perturbative modes
in the high temperature phase seems to be a real possibility, as also
the existence of some colour degrees of freedom till lower temperatures.
These issues deserve urgent attention in the context of QGP
diagnostics.
\par
The work of SKG was supported in part by the Council of Scientific \& Industrial 
Research, Govt. of India. 

\end{document}